\documentclass[12pt]{article} 
\usepackage{graphicx}
\usepackage{amsmath,amssymb}
\usepackage{longtable}

\begin{document}
\baselineskip 21pt
\setcounter{totalnumber}{8}

\bigskip

\centerline{\Large \bf Thickness of Stellar Disks in Early-type Galaxies}

\bigskip

\centerline{\large E. M. Chudakova$^1$ and O.K. Sil'chenko$^1$}

\noindent
{\it Sternberg Astronomical Institute of the Lomonosov Moscow State University, Moscow, Russia}$^1$

\vspace{2mm}
\sloppypar 
\vspace{2mm}

\bigskip

{\small 
\noindent
We suggest and verify a new photometric method enabling derivation of relative 
thickness of a galactic disk from two-dimensional surface-brightness distribution of the galaxy 
in the plane of the sky. The method is applied to images of 45 early-type (S0--Sb) galaxies with known radial 
exponential or double-exponential (with a flatter outer profile) surface-brightness distributions. The 
data in the $r$-band have been retrieved from the SDSS archive. Statistics of the estimated relative 
thicknesses of the stellar disks of early-type disk galaxies show the following features. The disks of 
lenticular and early-type spiral galaxies have similar thicknesses. The presence of a bar results in only 
a slight marginal increase of the thickness. However, we have found a substantial difference between the thicknesses 
of the disks with a single-scaled exponential brightness profile and the disks that represent the inner segments 
of the Type III (antitruncated) profiles. The disks are significantly thicker in the former subsample than 
in the latter one. This may provide evidence for a surface-brightness distribution of a single-scaled exponential 
disk  to be formed due to viscosity effects acting over the entire period of star formation in the disk.
}

\clearpage

\section{INTRODUCTION}

Two main large-scale components of structure of any galaxy -- a stellar bulge and a stellar disk -- have different 
spatial geometries. Bulges are spheroids, with three axes of typically comparable length; while disks are flat 
stellar subsystems whose thicknesses are much smaller than their radii. However, the thickness of a stellar disk
is not infinitely small, and the thickness and the radius of a typical galactic disk differ by less than an 
order of magnitude. The thickness of a stellar disk is an important physical parameter which can help to specify 
dynamical evolution of the galaxy: interactions with other galaxies such as gravitational tides or merging with 
smaller satellites, can ``heat'' the disks (e.g. \cite{miwanoguchi,walker96}), making them thicker, while smooth, 
laminar accretion of a large amount of cool gas with subsequent star formation and the development of young stellar 
systems that are dynamically cool like the gas from which they form, can reduce the thickness of the disk 
\cite{reshcombes}.

Observational studies of the thicknesses of stellar disks in galaxies have been based to date mainly on statistics
of large inhomogeneous samples of galaxies. It was also assumed that the disks were characterized by random orientation
in space, so that a distribution of apparent ellipticities of the disks seen in projection onto the plane of
the sky carried information about the average ratio of the thickness to the radius. The latest statistical
estimates of the average thickness of galactic disks $q$ are based on the data from the SDSS: Padilla and
Strauss \cite{padilla_strauss} have analyzed a sample of several thousand galaxies with exponential brightness 
profiles -- i.e., galaxies dominated by disks -- using the Sixth Data Release of the SDSS, and they have obtained 
$q = 0.21 \pm 0.02$. The same team has obtained a somewhat higher estimate of $q$, $q = 0.267 \pm 0.009$, by using 
naked-eye classifications of spiral galaxies from the Galaxy Zoo project, by selecting them from the 8th Data Release 
of the SDSS \cite{padillanew}. There are not very many individual, non-statistical estimates of the thicknesses of 
stellar disks. It is possible to measure the thickness of a stellar disk directly only if the galaxy is viewed edge-on. 
Let us note that it is difficult to estimate other properties of the disk structure in this case, since all the material 
that is located in the equatorial plane is seen as integrated along the line of sight. In early studies of the 
surface-brightness distributions in the optical $BVRI$-bands in eight S0--Sd galaxies viewed edge-on \cite{degrijs},
very thin disks were found: the ratio of the vertical and horizontal scalelengths for the brightness distributions were, 
on average, $z_0/h = 0.17 \pm 0.01$. However, this might be due to the small size of the sample. Recently, based on a 
decomposition of the near-IR images of some two hundred edge-on galaxies ($JHK$-band images from the 2MASS survey), 
Mosenkov et al. \cite{mosenkov} found a very broad distribution of $z_0/h$ with a median value
close to 0.3. Meantime, the average thickness of galactic disks has no more meanings than the mean temperature in
a hospital. Galactic disks can suffer very different dynamical evolution, due mainly to the density
of their environment, as well as to the prominence and structure of the spheroidal components of the galaxy.
Correspondingly, the distribution of disk thicknesses is far from a normal law. Estimate of individual
disk thicknesses and their comparison with the statistics of other disk parameters, such as the surface-brightness
profile shape, is much more interesting problem. This is the topic addressed in the present study. We propose here 
a new method, which we have used to estimate individual disk thicknesses in a sample of galaxies viewed at angles 
different from 90 degree. We have been interested in possible differences of disk thicknesses in galaxies with different 
radial surface-brightness profiles. We used the sample of galaxies and classification of the radial brightness profiles
compiled in the papers by Erwin et al. \cite{erwin08} and Guti\'errez et al. \cite{erwin11}. In contrast to the classical 
study by Freeman \cite{freeman}, which reported strictly exponential surface-brightness distributions in the stellar 
galactic disks, Erwin with coauthors \cite{erwin08,erwin11} found that, in general, a disk surface-brightness profile
can be represented by two exponential segments with different characteristic scalelengths. It is currently
believed that single-exponential disks with a uniform scale (Type I), truncated disks (outer scalelength smaller
than the inner one, Type II), and two-tiered disks (outer scalelength larger than the inner one, Type III) have
different origins. Previously proposed scenarios of their formation predicted definite consequences for the disk 
thicknesses. Thus, our results can be used to test theories of the formation and evolution of the galactic disks.

\section{The Method}

We have developped an original method which can be used to estimate relative disk thicknesses by using
two-dimensional surface-brightness distributions of a galactic disk projected onto the plane of the sky. 
The method is applicable to galactic disks if (a) the disks are circular and (b) with exponential or 
double-exponential surface-brightness profiles. The disks are considered to be plane-parallel (within 
every exponential segment of the brightness profile), i.e., they have the same thickness at any distance 
from the center. We have assumed that the same exponential law for the fall of the volumetric
stellar emissivity with radius is obeyed in the equatorial plane and at some distance from this plane. 
For the method to work correctly, it is also necessary that the galaxy not be viewed strictly face-on or 
strictly edge-on. The key idea of our approach is analysis of the surface-brightness distribution in the
polar coordinate system and splitting the galaxy image into sectors. The azimuthal tracing of equal surface
brightness (isophotes) and measurements of the exponential scalelength of the radial brightness
profiles at different angles to the major axis of the isophotes provide discrete sets of points 
separated by equal angles in polar coordinates referenced to the center of the galaxy and its line 
of nodes. Both the isophotes and azimuthal changes of the exponential scalelengths were fitted by ellipses. 
This approach makes it possible to estimate independently the axis ratios of the isophotes and of the 
ellipses for the scalelengths; we find the relative disk thickness by comparing these two ellipticities.

\subsection{The Main Formulae}

\begin{figure*} 
\includegraphics[width=0.5\hsize]{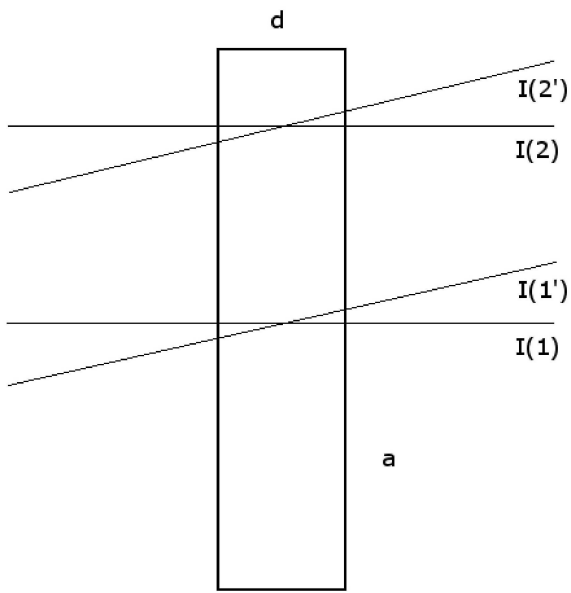} 
\caption{Sketch showing the integration of the disk stellar emissivity along the line of sight 
for various disk inclinations to the plane of sky. The disk is shown edge-on, and the observer's 
line of sight lies in the plane of the picture.} 
\end{figure*}

Let us consider a galaxy with a round disk of radius $a$, with relative thickness $q = d/a$ and inclination 
to the plane of the sky $i$. The ellipticity of the isophotes $e_I$ of the image of this disk is related 
to its thickness and inclination by the formula:

\begin{equation}
\sin i = \sqrt{\frac{2e_I - e_I ^2}{1-q^2}}
\label{formiso}
\end{equation}

Let us assume that the disk surface brightness profile has an exponential shape, being expressed
in the plane of a galaxy by a formula $I(R) = I(0) \exp(-R/h)$ \cite{freeman}. Let us compare the 
surface brightnesses near the points 1 and 2 (Fig.~1) for viewing along the rotation axis of the disk 
($I(1)$ and $I(2)$) and at some angle to this axis ($I(1')$ and $I(2')$). We assume that the disk 
thickness $d$ is constant with radius and that the radial scalelength $h$ does not depend on the 
distance to the galactic plane (on the height). The surface brightness seen at a given point of the disk 
image is the integral of the volumetric emissivity along the line of sight. For the lines of sight 
corresponding to $I(1)$ and $I(2)$, the volumetric emissivity in each element of the parallel segments 
of the lines of sight over which we are integrating have the same ratio: $I(1)/I(2) = \exp (-\rho_{12}/h)$. 
If $d$ is constant along the radius and $h$ is constant with height above the equatorial plane, 
the proportionality is retained in each element of the lines of sight corresponding to $I(1')$ and
$I(2')$. Therefore, $I(1')/I(2') = \exp(-\rho _{12}/h)$. Thus, the scalelength $h$ measured along the line 
of nodes does not depend on the inclination of the line of sight to the galactic plane. Then, 
the scalelength $h$ measured in the plane of sky at an angle to the line of nodes would decrease entirely 
due to the effect of projection, obeying to a cosine dependence on the polar angle. Thus, having derived 
an ellipse for the distribution of the apparent exponential scalelengths over the azimuth in the plane of the sky 
and after measuring the ellipticity of this ellipse, $e_h$, we can find the inclination of the galactic disk 
to the plane of sky, $i$, from the following relation:

\begin{equation}
\sin i =\sqrt{2e_h - e_h ^2}.
\label{formscale}
\end{equation} 

Thus, substituting (\ref{formscale}) into (\ref{formiso}), we obtain the following work formula for deriving 
the thickness of a galactic disk from the ellipticities of the isophotes and of the azimuthal distribution 
of the exponential scalelength:

\begin{equation}
q = \sqrt{1 - \frac{2e_I - e_I ^2}{2e_h - e_h ^2}}.
\label{finform}
\end{equation}

\subsection{Steps of the Computations}

Prior to the disk thickness calculations, the sky background was subtracted from the galaxy images, 
and foreground stars and other bright, compact features such as rings and starforming regions were masked.

An isophote radial position $R_n(count)$ can be determined at different polar angles from the equation 
$I_n(R_n)=count$ for a fixed signal level of $count$. An ellipse with the semi-major axis $a_I$ and 
ellipticity $e_I$ fitted to a set of 20 $R_n(count)$ points uniformly distributed over azimuth in polar 
coordinates, $M_n(count)=(R_n, \frac{\pi n}{10}+\frac{\pi}{20})$, describes the galaxy isophote at the 
brightness level $count$. We have divided the cleaned galaxy image into 20 sectors by the rays started from
the galaxy center. We took the center at the brightest point of the image, and divided the image into 
20 nonintersecting sectors with opening angle $\pi /10$. We constructed radial surface-brightness distributions 
$I'_n (\rho)$ for every $n$th sector as the dependencies of the brightness $I'$ averaged along the
azimuth on the distance $\rho$ from the center, by averaging $I'$ over the azimuthal angles 
$\varphi \in \left( \frac{\pi n}{10} ; \frac{\pi (n+1)}{10}\right]$. 
If the galaxy possesses central symmetry, the profiles $I' _n(\rho)$ and
$I'_{n+10}(\rho)$ are identical, and it makes sense to increase the accuracy of measurements by considering 
only 10 profiles, $I_n(\rho)=(I'_n(\rho)+I'_{n+10}(\rho))/2$. Assuming an exponential shape of the radial 
dependencies for the disk surface brightness, we calculated the scalelength $h_n$ for each of the 10 sectoral 
cross-sections.  For this purpose, we fitted the straight line $1.086(R_m/h) + l$ to
the set of points $(R_m,2.5\log I_n(R_m))$, where $R_m \in [R_{n0}(j),R_n(count)]$ by the least-square method. 
We have chosen the $R_{n0}$ such as those with the smallest least-square scatter of the surface brightness
measurements around the exponential law, individually for each cross-section.
Then, $h_n=h(R_{n0})$ is the scale for the $n$-th cross-section in the radial range $[R_{n0}(count),R_n(count)]$. 
The set of points in polar coordinates, representing exponential scalelengths in the different sectors,
$N_n=(h_n, \frac{\pi n}{10}+\frac{\pi}{20})$
is approximated by an ellipse with the semi-major axis $a_h$ and ellipticity $e_h$.
For every Type III galaxy, for which the full disk surface-brightness profile is approximated by two
exponential segments, we constructed the isophotes and ellipses of the scalelength coefficients for each 
of the two exponential segments separately. The thickness of the disks was derived using (\ref{finform}).

\section{The galaxy sample and photometric data used.}

To test the method, we have taken a sample of galaxies with exponential and double-exponential
disks which radial brightness profiles have been analyzed by Erwin et al.\cite{erwin08} and Gutierrez et al.\cite{erwin11}. 
The galaxies with bars were described in \cite{erwin08}, and the galaxies without bars --
in \cite{erwin11}. We have selected galaxies with single-scaled exponential disks (Type I) and with 
double-scaled exponential disks, where the scalelength of the outer disk is larger than the scalelength
for the inner disk (Type III), from the total sample presented in \cite{erwin08,erwin11}. 
Our final sample includes 66 galaxies: 29 galaxies of Type I and 37 galaxies of Type III. The sample
contains mainly early-type (S0--Sb) disk galaxies. We have retrieved digital images of the sample galaxies 
from the database of the SDSS photometric survey, Data Release 8 \cite{sdssdr8}. The images were 
found for 24 Type I and 19 Type III galaxies. We have also adopted archival SCORPIO data \cite{scorpio}
from the 6-m telescope of the Special Astrophysical Observatory for NGC~2300 and NGC~2787. 
We have applied our method to the SDSS $r$-band images, which have the highest signal-to-noise ratio. 
It have appeared impossible to apply the method to NGC~2712 (Type I; the disk looks non-exponential), 
NGC~4045 (Type III; the inner disk is non-exponential, and the outer disk looks noisy), NGC~5806 
(Type III; the inner disk displays a high intrinsic ellipticity), UGC~4599 (Type III; low signal-to-noise 
ratio has prevented to extract the low-surface-brightness disk). We have also added to our sample the galaxies
NGC~4513 (Type III, was analyzed in \cite{uvrings}) and NGC~4124 (Type I, was inspected in \cite{n4124}).
As a result, we have succeeded to apply our method to derive the relative disk thicknesses 
for 26 Type I galaxies, 17 inner disks of Type III galaxies, and 2 outer disks of Type III galaxies.

\begin{figure*} 
\includegraphics[width=\hsize]{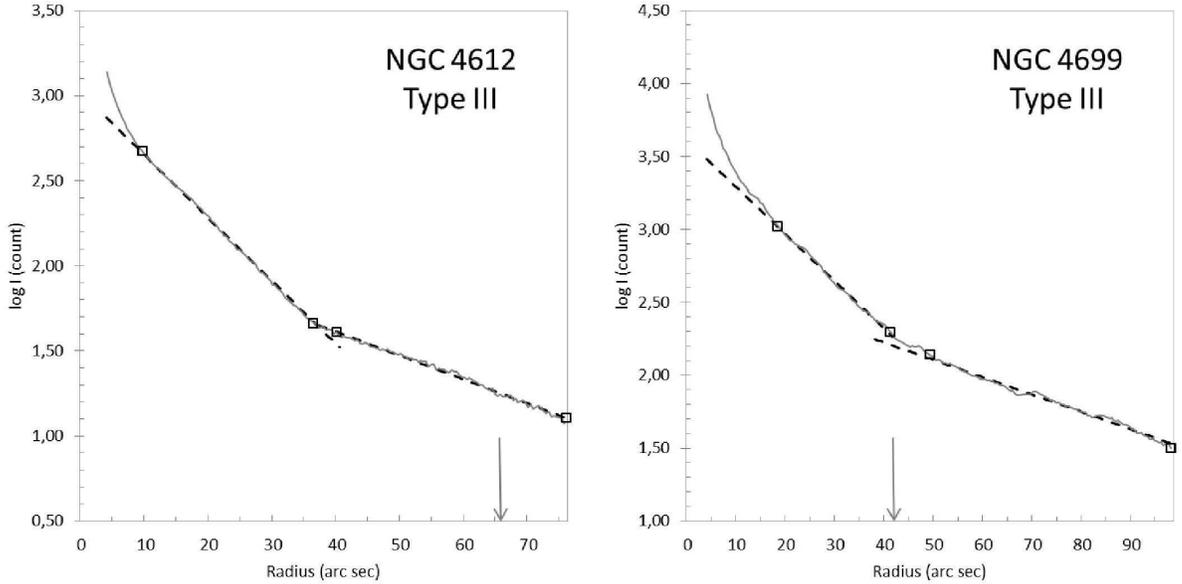} 
\caption{Double-exponential surface-brightness profiles of the 
Type III galaxies NGC 4612 and NGC 4699 in the SDSS $r$-filter. The arrows mark the $R_{25}$-radii.} 
\end{figure*}

\section{RESULTS}

The accuracy of the SDSS data has appeared to be sufficient to derive the thicknesses of both 
the inner and outer disks for only two Type III galaxies: NGC~4612 and NGC~4699. Figure~2 presents 
the surface brightness profiles in the sectoral cross-sections along the major axes of the 
inner disks of these galaxies. The arrow marks the isophotal radius at the level of $\mu _r =23.5$, 
that corresponds roughly to the optical galactic radius $R_{25}$\, where we have aimed to draw the reference 
isophotes. The outer disk of NGC~4612 is probably the main stellar disk, with the central surface
brightness typical for disk galaxies \cite{freeman}, while the outer disk in NGC~4699 is traced beyond 
the formal optical boundary of the galaxy and has low central surface brightness. Nevertheless, 
the outer-disk parameters in these galaxies are surprisingly similar: in both cases, the outer disks 
are thick ($q =0.52-0.56$), being much thicker than their inner disks ($q =0.19-0.26$). Since we 
already found a similar trend of the disk flaring while going outward along the radius, 
by measuring the inner and outer disk thicknesses of the Type III galaxy NGC~7217 from kinematic 
data \cite{n7217last}, this result hints at some tendency, with the outer disks most often 
being thicker than the inner ones. 

\centerline{Table 1: Relative thicknesses of the stellar disks,\\}
\centerline{found by our new photometric method in the galaxies of our sample.}
{\small
\begin{longtable}{|r|p{3.5cm}|r|c|r|r|r|l|l|}
\hline
 Galaxy & Type(NED) & $R_{25}^{\prime \prime}$ & Disk type & $h^{\prime \prime}$	& $R_0^{\prime \prime}$
& $R_I^{\prime \prime}$ & $q$ & $\delta q$	\endhead
\hline
IC 676 	&	(R)SB(r)0+	&	40	&	I-1	&	72	&	-	&	61	&	0.3790	& 0.0001 \\
NGC 1022	&	(R')SB(s)a	&	61	&	I-1	&	56	&	-	&	97	&	0.808	&	0.012	\\
NGC 2300	&	 SA0$^0$  	&	85	&	I-2	&	416	&	-	&	187	&	0.565	&	0.023	\\
NGC 2787 	&	 SB(r)0+ 	&	95	&	I-1	&	85	&	-	&	114	&	0.230	&	0.008	\\
NGC 3032	&	 SAB(r)$0^0$	&	36	&	I-2	&	36	&	-	&	36	&	0.415	&	0.016	\\
NGC 3169	&	SA(s)a	&	125	&	I-2	&	126	&	-	&	125	&	0.40	&	0.05	\\
NGC 3485 	&	 SB(r)b	&	47	&	I-1	&	92	&	-	&	47	&	0.70	&	0.02	\\
NGC 3489	&	SAB0$^+$(rs)	&	87	&	III-d-1	&	56	&	54	&	111	&	0.0	&	0.03	\\
NGC 3599	&	 SA0	&	16	&	I-2	&	34	&	-	&	30	&	0.706	&	0.007	\\
NGC 3604	&	SA(s)a pec	&	41	&	III-d-2?	&	45	&	28	&	55	&	0.15	&	0.01	\\
NGC 3607	&	SA(s)$0^0$	&	114	&	I-2	&	103	&	-	&	114	&	0.31	&	0.11	\\
NGC 3619	&	(R)SA$0^-$(s):	&	81	&	III-d-2	&	82	&	42	&	82	& 0.22	& 0.03	\\
NGC 3626	&	(R)SA(rs)0+	&	71	&	I-2	&	64	&	-	&	49	&	0.52	&	0.02	\\
NGC 3898	&	SA(s)ab	&	94	&	III-d-2	&	77	&	41	&	81	&	0.175	&	0.007	\\
NGC 3900	&	SA0$^+$(r)	&	67	&	III-d-2	&	70	&	50	&	100	&	0.195	&	0.008	\\
NGC 3998	&	SA0$^0$(r)?	&	71	&	III-d-2	&	57	&	57	&	113	&	0.35	&	0.03	\\
NGC 4124	&	SA0$^+$(r)	&	100	&	I-1	&	74	&	-	&	100	&	0.0	&	0.007	\\
NGC 4138	&	SA0$^+$(r)	&	65	&	III-d-2	&	39	&	19	&	46	&	0.066	&	0.008	\\
NGC 4150	&	SA0$^0$(r)?	&	45	&	III-s-2	&	30	&	32	&	68	&	0.0	&	0.004	\\
NGC 4151 	&	(R')SAB(rs)ab	&	90	&	I-1	&	90	&	-	& 90	&	0.45	&	0.04	\\
NGC 4245 	&	SB(r)0/a	&	65	&	I-1	&	78	&	-	&	88	&	0.687	&	0.006	\\
NGC 4267 	&	SB(s)0-?	&	74	&	I-1	&	76	&	-	&	74	&	0.48	&	0.01	\\
NGC 4340	&	 SB(r)0+   	&	74	&	I-1	&	149	&	-	&	74	&	0.524	&	0.009	\\
NGC 4371	&	SB0$^+$(r)	&	113	&	III-d-1	&	154	&	126	&	149	&	0.362	& 0.008	\\
NGC 4459	&	SA0$^+$(r)	&	32	&	III-d-2	&	38	&	16	&	45	&	0.524	&	0.024	\\
NGC 4477	&	 SB(s)0:? 	&	114	&	I-1	&	85	&	-	&	88	&	0.69	&	0.03	\\
NGC 4513	&	(R)SA$0^0$	&	26	&	III-1	&	28	&	14	&	26	&	0.245	&	0.004	\\
NGC 4578	&	SA(r)$0^0$	&	72	&	I-2	&	79	&	-	&	72	&	0.0	&	0.02	\\
NGC 4596 	&	SB(r)0+	&	182	&	I-1	&	174	&	-	&	102	&	0.664	&	0.014	\\
NGC 4612	&	(R)SAB$0^0$	&	68	&	III-d-1 (exter)	&	98	&	44	&	88	& 0.516 & 0.020	\\
NGC 4612	&	(R)SAB$0^0$	&	68	&	III-d-1 (inter)	&	38	&	10	&	40	& 0.192	& 0.006	\\
NGC 4643 	&	SB(rs)0/a	&	105	&	I-1	&	182	&	-	&	104	&	0.73	&	0.02	\\
NGC 4665 	&	SB(s)0/a	&	107	&	I-1	&	145	&	-	&	106	&	0.812	&	0.007	\\
NGC 4691	&	(R)SB(s)0/a pec	&	71	&	III-d-1	&	105	&	32	&	65	& 0.715	& 0.013	\\
NGC 4699	&	SAB(rs)b	&	130	&	III-d-1 (exter)	&	98	&	49	&	98	& 0.564	& 0.007	\\
NGC 4699	&	SAB(rs)b	&	130	&	III-d-1 (inter)	&	34	&	18	&	42	& 0.258	& 0.013	\\
NGC 4754 	&	SB(r)0-: 	&	114	&	I-1	&	129	&   -	  	&	113	&	0.34	&	0.02	\\
NGC 4772	&	SA(s)a	&	107	&	I-2	&	116	&	-	&	106	&	0.31	&	0.02	\\
NGC 5485	&	SA0 pec     &	58	&	I-2	&	55	&	-	&	58	&	0.43	&	0.02	\\
NGC 5520	&	Sb	&	37	&	I-2	&	38	&	-	&	36	&	0.214	& 0.014	\\
NGC 5740	&	SAB(rs)b	&	61	&	III-d-1	&	46	&	27	&	69	& 0.0	&	0.018	\\
NGC 5750 	&	SB(r)0/a	&	70	&	I-1	&	42	&	-	&	70	&	0.0	&	0.012	\\
NGC 7177	&	SAB(r)b	&	53	&	III-d-1	&	42	&	34	&	49	&	0.16	&	0.04	\\
NGC 7457	&	SA0$^-$(rs)?	&	92	&	III-d-2	&	48	&	12	&	27	&	0.20	&	0.02	\\
NGC 7743 	&	(R)SB(s)0+	&	57	&	I-1	&	68	&	-	&	75	&	0.50	&	0.04	\\
\hline
\end{longtable}
%\end{table*}
} 

The results of disk thickness measurements for our total sample are presented in the Table, 
whose columns contain (1) the galaxy name, (2) the morphological type according to the 
NED database, (3) the semi-major axis of the galaxy optical image according to the NED data, 
(4) the type of the surface brightness profile over the disk according to \cite{erwin08,erwin11},
(5) the exponential scalelength along the disk line of nodes according to our results, 
(6) the assumed inner boundary of the disk in Type III galaxies, (7) the major-axis radius at which the reference 
isophote was drawn, (8) the relative disk thickness, $q\equiv z_0/h$, according to our results, and (9) the 
statistical uncertainty of the relative disk thickness $q$. 

Next several plots show the statistics of our results on the relative thicknesses of the stellar disks, 
presented as histograms. The abscissa traces the disk thickness $q$, and the column heights correspond
to the number of disks with a specified thickness in the subsamples. Galaxies for which the ellipticities 
of the isophotes and scalelength projection distributions are close enough so that the argument
of the square root in (\ref{finform}) may be formally negative due to the statistical uncertainty in $q$, 
are plotted against negative $q$\, values. We take such disks to be infinitely thin, but plot them to the left 
from zero, in the ``complex'' range.

\begin{figure*} 
\includegraphics[width=\hsize]{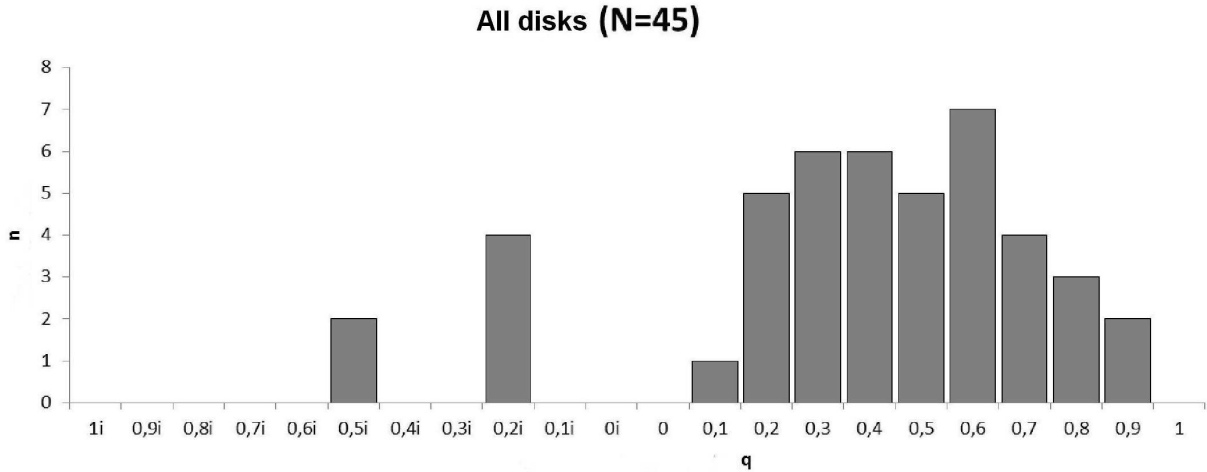} 
\caption{The distributions of the relative thicknesses $q$ for all the stellar disks analyzed. The ``complex'' region 
to the left contains objects that have formally negative arguments of the square root in (\ref{finform}); these are 
either very thin disks with $q\approx 0$, which ended up in the $0.2i$-bin due to the statistical uncertainty in $q^2$, 
or disks whose intrinsic shapes suffer notable non-axisymmetry -- two disks in the bin 0.5$i$ can be probably attributed to 
those.} 
\end{figure*} 

\begin{figure*} 
\includegraphics[width=\hsize]{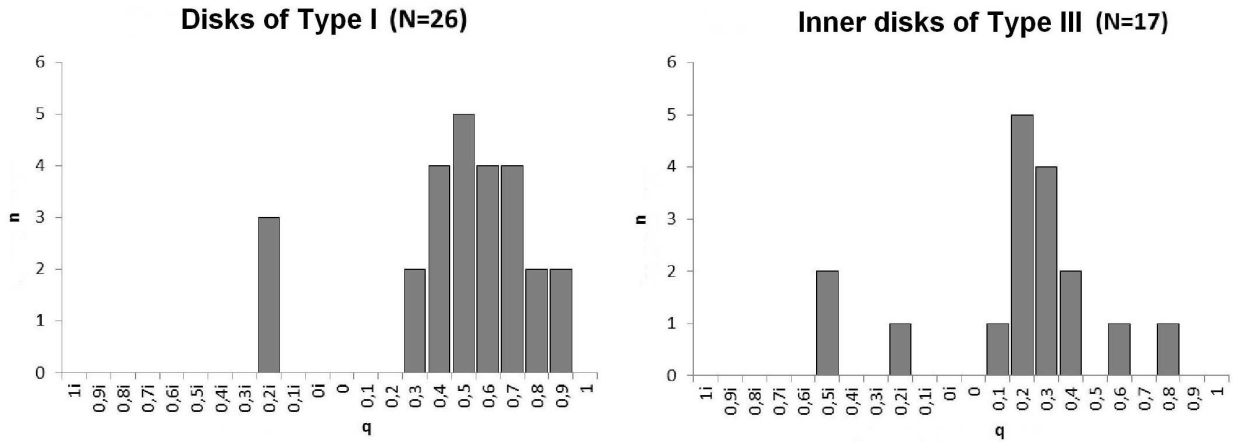} 
\caption{Comparison of the relative disk thicknesses in the Type I galaxies (single-exponential disks) 
and in the inner disks of the Type III galaxies (two-tiered double-exponential disks).} 
\end{figure*}

At the histogram for the whole sample (Fig.~3), the mean disk thickness is 0.38. The histogram for 
the whole sample is very broad, with relative thicknesses ranging from 0 to 0.9. If we divide
the whole sample into two subsamples according to the profile type of the disks, the distributions
for the subsamples appear to be much narrower. 
Figure~4 shows that the distribution of the inner-disk thicknesses for the Type III galaxies 
and that of the disk thickness for the Type I galaxies have shifted maxima. The inner 
disks of the Type III galaxies have a mean thickness of 0.19, while the disks of the Type I galaxies 
have a mean thickness of 0.48. Despite the small sizes of the subsamples, the Kolmogorov-–Smirnov test 
indicates that the probability that the disks of Type I galaxies and the inner disks of Type III galaxies 
belong to the same parent sample is only 0.7\%; i.e., the relative disk thicknesses for Type I and Type III galaxies 
are significantly different, with Type I galaxies (exponential disks with a unique scalelength) having thicker disks.

\begin{figure*} 
\includegraphics[width=\hsize]{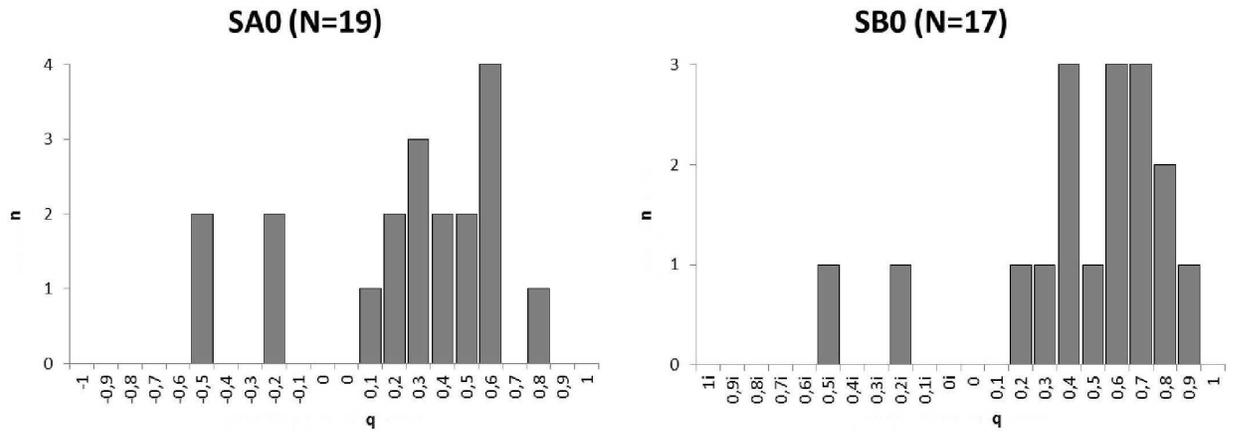} 
\caption{Comparison of the distributions of the relative disk thicknesses in lenticular galaxies with and without bars --
SB0 versus SA0.} 
\end{figure*}

\begin{figure*} 
\includegraphics[width=\hsize]{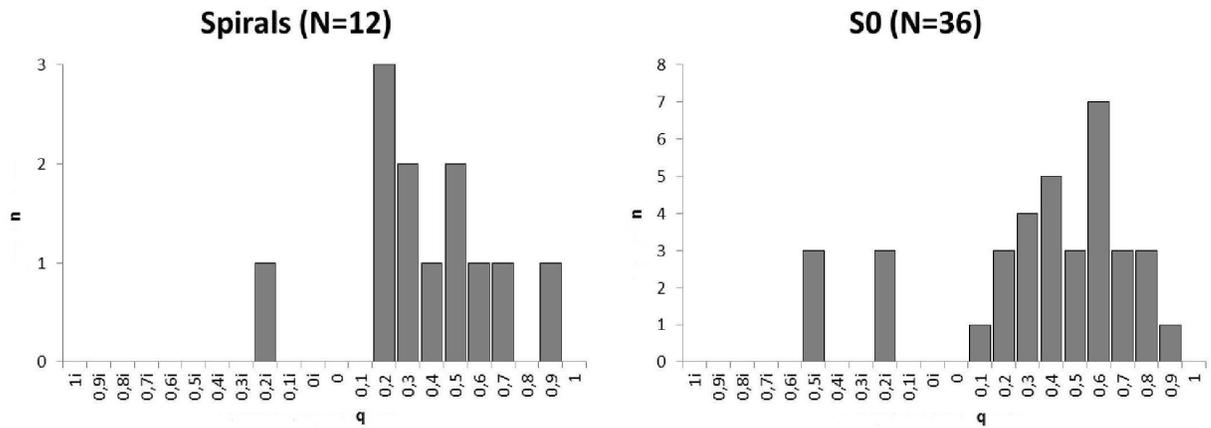} 
\caption{Comparison of the distributions of the relative disk thicknesses in lenticular and early-type spiral (Sa--Sb) galaxies.} 
\end{figure*}

It is interesting that if the galaxies are separated into lenticular and spiral galaxies, or into galaxies with or 
without bars, the difference in the disk thicknesses between the subsamples is not so significant. For example, 
the mean disk thickness for the lenticular galaxies without bars is 0.26, while the lenticular galaxies with bars 
have a mean disk thicknesses of 0.46 (Fig.~5). However, both subsamples show a large scatter of the thicknesses. 
The Kolmogorov--Smirnov test indicates that the probability for the galaxies with and without bars to belong to the same 
parent distribution in terms of their disk thickness is 38\%. If we consider the SA0 and SB0 galaxies as one subgroup (S0) 
and compare them to the early-type spiral galaxies, the significance of the difference becomes even lower. The mean 
relative disk thickness for all the lenticular galaxies is 0.36, and for the spiral galaxies 0.38 (Fig.~6). The Kolmogorov-–Smirnov 
test indicates that the probability that these histograms are drawn from the same parent distribution is 99.5\%. 

\section{THE COMPARISON OF OUR RESULTS WITH EARLIER STATISTICS}

\begin{figure*} 
\includegraphics[width=\hsize]{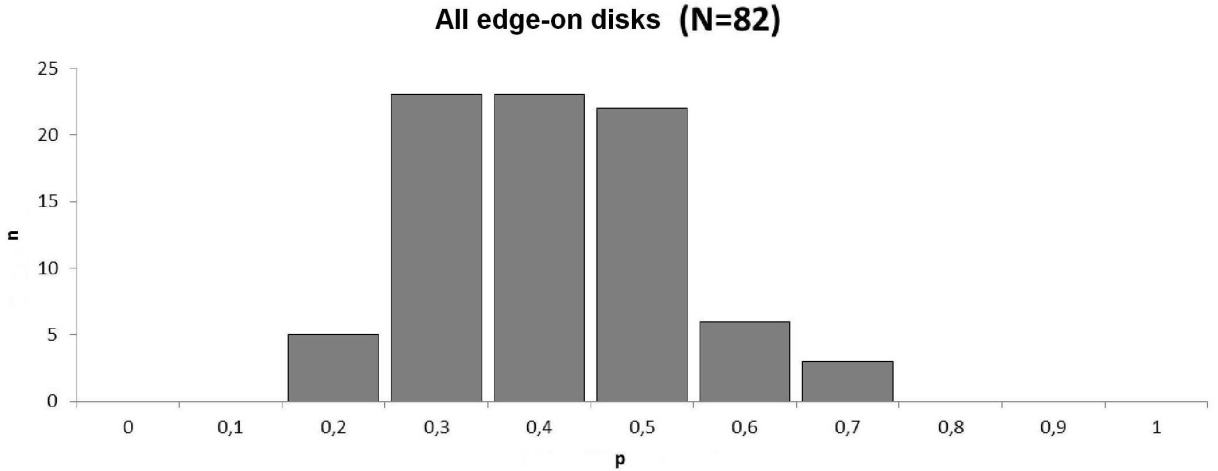} 
\caption{The distribution of the relative disk thicknesses for the edge-on galaxies from the sample by
Mosenkov et al. \cite{mosenkov} (the morphological types S0--Sb and the parameters in the NIR $J$--band are taken).} 
\end{figure*}

We have compared our results with some earlier statistics concerning the sample of edge-on galaxies compiled 
by Mosenkov et al.\cite{mosenkov}, who considered a sample of galaxies from the 2MASS survey which were selected
according to following criteria. 
\begin{enumerate}
\item{The galaxies are seen strictly edge-on, $i=90^{\circ}$.} 
\item{The axis ratio of the image in the co-added $J + H + K$\, frame exceeds 0.2.} 
\item{The radius of the galaxy measured in the $K$-band $R_{Kron} > 30^{\prime\prime}$.} 
\item{The galaxies seem to be non-interacting.} 
\item{The concentration index exceeds 2.0.}
\end{enumerate}
This approach had provided a sample of 175 disk galaxies of all types, whose $JHK$-images were decomposed 
into disks and bulges by using an original algorithm. Along the radius, the surface brightness profile was 
fitted by an exponential law with a single scalelength $h$, and the vertical decrease of the surface brightness 
from the equator outside was approximated by a squared secant law with a scale $z_0$. The ratio of our parameters 
$a$ and $d$ and the ratio of the scalelengths $h$ and $z_0$ are different characteristics. Nevertheless, it is 
interesting to compare the distributions of $d/a$ and $z_0/h$, which are in fact various estimators of the 
the relative thicknesses of the galactic disks. We have extracted early-type disk galaxies (S0--Sb) from 
the whole sample \cite{mosenkov}, and have calculated $p = z_0/h$ by using quantitative estimates 
of the disk parameters kindly provided to us by Dr. N.Ya. Sotnikova in the tabular form. We are presenting here
the  histograms of the $p$ distributions for the resulting subsample of 82 early-type disk galaxies. The mean 
relative thickness over the full subsample is 0.41 (Fig.~7); the scatter of these values around the mean is not 
too large --  about 83\%\ lie between 0.3 and 0.5. Nevertheless, although the mean disk thicknesses for our sample 
and for the sample \cite{mosenkov} are similar, the Kolmogorov--Smirnov test indicates that the probability 
that our total sample and the subsample from \cite{mosenkov} belong to the same parent sample is only 1.7\%.
The mean disk thickness for the subsample \cite{mosenkov} is the most close to our subsample
of Type I disks, but they are also statistically distinguishable. This may be associated with the small
size of our sample of Type I galaxies. Dividing the subsample \cite{mosenkov} into two ones according 
to the morphological types of galaxies does not result in appreciable changes in the character of their
distribution, as it has been shown for our own subsamples. The mean disk thicknesses of lenticular and edge-on
spiral galaxies are not significantly different: 0.40 versus 0.42 (Fig.~8). This is also in qualitative 
agreement with our results. However, the histogram widths for these two galaxy subsamples are substantially
different: only 76\%\ of lenticular galaxies have $p$-values between 0.3 to 0.5, while for the spiral galaxies
the distribution of the thicknesses is more narrow: 89\%\ of them have $p$-values in the range [0.3, 0.5]. 
The Kolmogorov--Smirnov test indicates an absence of significant differences between
the disk thicknesses for galaxies of various morphological types: the probability that both subsamples
belong to the same parent distribution is 76\%. Thus, our sample and the sample \cite{mosenkov} 
are similar in the sense that, among early-type disk galaxies, there is no significant difference in the disk 
thicknesses of lenticular and spiral galaxies.

\begin{figure*} 
\includegraphics[width=\hsize]{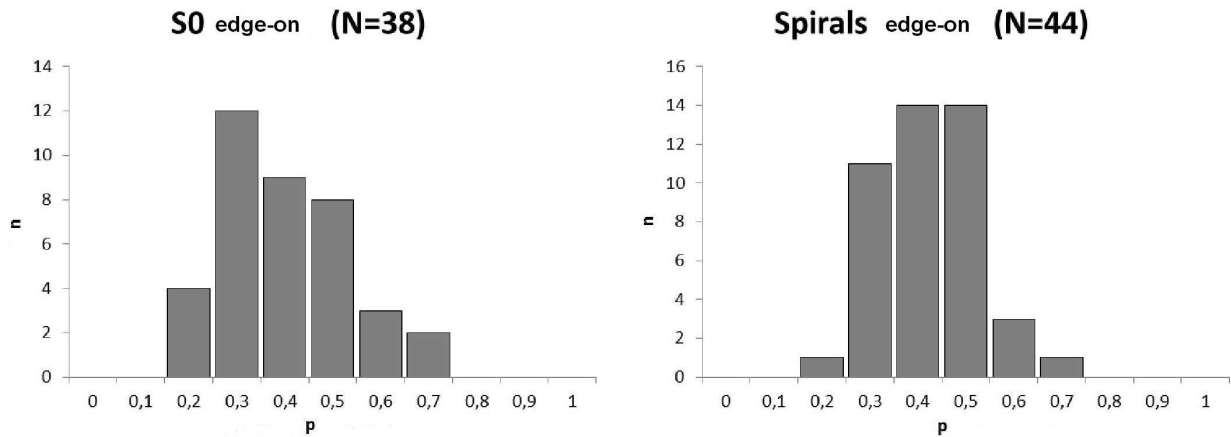} 
\caption{The comparison of the distributions of relative disk thicknesses for edge-on lenticular and spiral 
(Sa--Sb) galaxies by analyzing the decomposition results of their J-band images from \cite{mosenkov}.} 
\end{figure*}

\section{CONCLUSIONS AND DISCUSSION}

In this work we propose a new photometric method to derive relative thicknesses of galactic disks 
from their two-dimensional surface-brightness distributions projected onto the plane of the sky. 
We have applied this method to 45 early-type (S--Sb) disk galaxies with purely exponential
(Type I according to \cite{erwin08,erwin11}) or double-exponential,
with a flatter outer profile (Type III according to \cite{erwin08,erwin11}).
The obtained statistics on the relative disk thicknesses of early-type galaxies reveals a lot
of interesting features. On average, the stellar disks of lenticular and early-type spiral galaxies have 
similar thicknesses. A bar presence results in only marginal thickening of the disks. However, 
there is a significant difference between the thicknesses of the disks
with single-exponential brightness profiles and the inner parts of the disks with Type III profiles 
(i.e., with outer exponential disks having larger scalelengths). The disks are significantly thicker 
in the former than in the latter case.

How can we explain these similarities and differences in the frame of current ideas about
the formation and evolution of the stellar disks in galaxies? The nature of the exponential brightness 
(density) profiles of the stellar galactic disks remains still a mystery. In the recent observational 
study Hunter et al.\cite{hunteretal} noted that a single-scale exponential profile can be traced over the
whole galaxy, covering the areas with very different star-formation modes and timescales. 
This contradicts apparently the idea that an exponentially-shaped density profile develops gradually in 
the course of the disk's evolution, and provides some evidence for a scenario where primordial exponential 
density profile with a fixed scalelength is a necessary initial condition for the evolution of a stellar 
disk. However, if star formation proceeds through different modes and different timescales over different 
regions of a galaxy, why is the initial density profile not``smeared out'' over a timescale of a few Gyr 
through the disk secular evolution? 

A completely opposite paradigm developped analytically by Lin and Pringle \cite{linpringle} and later elaborated 
in numerical models in \cite{slyz} seems to be more physically motivated as concerning
the stellar disk thickness evolution as well. These studies considered a violent radial gas redistribution,
simultaneous with the star formation. If a gaseous galactic disk is hardly viscous and if
the viscosity timescale is comparable to the star formation timescale, an exponential profile of
the stellar disk is naturally formed. This model is not very popular, mainly because the viscosity is 
required to be too strong. However, at high redshifts, $z>2$, when relatively cool gas contributed 
a dominant fraction of baryonic matter in galaxies, and gaseous clumps with a range of sizes formed 
in the disks due to various instabilities and were suspended in more diffuse gas, the viscosity could be 
considerably stronger than it is known at the current epoch. 

Through investigation of the stellar populations at the centers of galaxies with single-exponential 
disk-brightness profiles (Type I), we received already some hints that radial gas inflow into the centers
of galaxies had to be especially intense in Type I disks. We found chemically distinct nuclei with abnormally 
large metallicity differences between the nuclei and the bulges in the Type I galaxies \cite{sauron_exp}.
If the Type I galaxies experienced the most violent radial gas redistribution during 
their secular evolution, it seems possible that the same processes that removed the angular momentum of 
the gas (transient bars of a tidal nature, for example) also heated dynamically the 
stellar disks. This could explain the presence of the thicker stellar disks in these galaxies. 

The origin of Type III disks is currently explained by minor mergers, following the dynamical simulations
by Younger et al.\cite{younger_sim}. Their models showed that outer exponential segments of the disk 
brightness profiles (which have larger exponential scalelengths in the case of Type III galaxies) are built
up during the event by old stars of the primary (``parent'') disk galaxy, which were undergoing 
violent outward migration due to dynamical tides from a satellite that impacted the disk. If the 
absorbed satellite contained also gas, it may concentrate in the central region of a galaxy and, 
after a subsequent star formation burst, adds a young, cool stellar population to the inner disk. 
This scenario can qualitatively explain the small thicknesses of the inner disks of the Type III galaxies 
and also the fact that, when we are able to investigate both segments of a Type III profile, the outer 
disks have often larger thicknesses than the inner disks.

\section{ACKNOWLEDGMENTS}

We thank N.Ya. Sotnikova for providing us with the results of the image decomposition for edge-on galaxies
from Mosenkov et al. (2010)\cite{mosenkov} in tabular form. We have constantly used the Lyon-Meudon extragalactic database 
(HYPERLEDA), which is supported by the LEDA team in the Lyon Observatory CRAL (France), and also the 
NASA/IPAC Extragalactic Database (NED), which is operated by the Jet Propulsion Laboratory, California 
Institute of Technology, under contract with the National Aeronautics and Space Administration. We have 
made use of the public archives of the SDSS-III survey (http://www.sdss3.org), which is supported by 
Alfred P. Sloan Foundation, the participant institutes of the SDSS collaboration, National Science Foundation,
and the United States Department of Energy.

\end{document}